\documentclass[letter, twocolumn]{jpsj2} 
\usepackage{graphicx}
\usepackage{latexsym} 

\usepackage{bm}
\newcommand{\en}{\varepsilon_n}

\newcommand{\ui}{{\rm i}} 
\newcommand{\blr}{{\bm r}}
\newcommand{\blq}{{\bm q}}

\title{
Anomalous Thermal Conductivity of Semi-Metallic 
  Superconductors with Electron-Hole Compensation
}

\author{Hiroto \textsc{ADACHI}
\thanks{E-mail address: adachi@itp.phys.ethz.ch} 
and Manfred \textsc{SIGRIST}}

\inst{Theoretische Physik, ETH-H\"{o}nggerberg, Zurich 8093, Switzerland}

\abst{
The effects of low carrier density and carrier compensation 
on mixed-state thermal transport are investigated 
beyond the quasiclassical approximation. 
It is shown that, contrary to the usual observations, 
the interplay of the two effects leads to an {\it increase} in the thermal 
conductivity immediately below the upper critical field $H_{c2}$. 
Our result can account for the anomalous behavior of the 
mixed-state thermal conductivity near $H_{c2}$ 
recently observed in URu$_2$Si$_2$. }

\kword{semi-metallic superconductor, electron-hole compensation, 
mixed-state thermal conductivity, heavy fermion superconductor} 

\begin{document}
\maketitle

The discovery of the Nernst effect in the normal state region 
of underdoped cuprates~\cite{Xu} has initiated renewed interest in 
magneto-transport phenomena in exotic superconductors. 
While a fluctuating vortex contribution may be a natural 
explanation for the unusual Nernst signal in cuprate, 
two other important aspects for magneto-transport phenomena 
were recently pointed out by Behnia and coworkers~\cite{Bel1,Behnia1}: 
the effects of electron-hole compensation~\cite{Bel1} 
and low carrier density~\cite{Behnia1}. 
The recent experiments on ultraclean URu$_2$Si$_2$ by Kasahara 
{\it et al.}~\cite{Kasahara} have motivated our study of these two points 
with regard to mixed-state thermal transport. 
The most intriguing result of ref.~\citen{Kasahara} is the 
{\it increase} in the low-temperature thermal conductivity 
below the upper critical field $H_{c2}$, 
showing a hump structure.~\cite{com2} 
Usually in the low-temperature limit, 
the thermal conductivity of a clean type-II superconductor 
{\it decreases} below $H_{c2}$~\cite{Maki}, due to
the enhancement of the Andreev scattering rate by vortices 
and the reduction of the density of states at the 
Fermi energy~\cite{Houghton,Adachi}. 

The key feature behind this anomalous behavior lies in the 
electronic structure of URu$_2$Si$_2$ introduced by the so-called
hidden-order phase occurring below $ T^*= 17.5$ K~\cite{Palstra,Maple,Baumann}. 
In this phase, the carrier density is drastically reduced, as seen in 
several transport measurements~\cite{Schoenes,Bel2,Behnia2}. 
Furthermore, a nearly perfect $H^2$-dependence of the magneto-resistance 
without any sign of saturation suggests the 
compensation of electron and hole pockets of the Fermi surface. 
This interpretation is further supported by the relatively small Hall angle. 

In this letter, we address the 
effects of the low carrier density and carrier compensation 
on thermal transport in the mixed phase using 
a simple model of a two-dimensional $s$-wave superconductor. 
This allows us to explain 
the unusual magnetic-field dependence of 
the thermal conductivity below $H_{c2}$ found in ref.~\citen{Kasahara} for 
URu$_2$Si$_2$, at least on a qualitative level. 
Since in this study it is necessary to go beyond the quasiclassical approximation, 
we generalize the approximation scheme for obtaining the Brandt-Pesch-Tewordt 
Green's function~\cite{Brandt} valid near $H_{c2}$, 
by employing the formalism of Vavilov and Mineev~\cite{Vavilov} 
originally developed to describe the mixed-state de Haas-van Alphen effect. 

We start by briefly reviewing the method of ref.~\citen{Vavilov}. 
Gor'kov equations for a two-dimensional $s$-wave superconductor 
under strong magnetic fields ($k_{\rm B}= c= \hbar= 1$) are as follows: 
\begin{equation}
  \big[\ui \en - \widehat{\cal H}_0(\blr) - \widehat{u}(\blr) 
  - \widehat{\Delta}(\blr) \big] \widehat{G}(\blr,\blr'; \ui \en) 
  = \delta(\blr-\blr'), 
  \label{Eq:Gorkov}
\end{equation}
where 
$\widehat{\cal H}_0 + \widehat{u} + \widehat{\Delta}  =  
( {{\cal H}_0 + u, \atop \Delta,}{ \Delta^* \atop -{\cal H}^*_0 - u} )$, 
$\widehat{G}= 
( {G, \atop F^\dag,} {F, \atop G^\dag} )$, 
and $\en = 2 \pi T (n+1/2)$ is the fermionic Matsubara frequency. 
The short-range impurity potential $u(\blr)$ obeys the Gaussian ensemble 
$\overline{u(\blr)}=0$; $\overline{u(\blr)u(\blr')}
= (1/m^*\tau) \delta(\blr- \blr')$ with $m^*$ and $\tau$ being 
the effective mass and the mean free time of quasiparticles, respectively. 
The single-particle Hamiltonian ${\cal H}_0$ is expressed as 
${\cal H}_0(\blr)= \frac{1}{2m^*}{\bm Q}^2 - \mu$, 
where ${\bm Q}= -\ui {\bm \nabla}+ |e|{\bm A}(\blr)$ 
with the vector potential ${\bm A}(\blr)$, and $\mu$ is the Fermi energy. 
For the magnetic field we assume the Landau gauge ${\bm A}(\blr)= Hx \widehat{\bm y}$, 
i.e., a uniform field as justified near $H_{c2}$, if the Ginzburg-Landau parameter 
$\kappa_{\rm GL}$ is large ( $\kappa_{\rm GL} \gg 1$ in URu$_2$Si$_2$~\cite{Ohkuni}). 
For simplicity, we have dropped the Zeeman coupling term, 
although paramagnetic effects  may be 
non-negligible in URu$_2$Si$_2$~\cite{Brison}. 
The paramagnetic limiting effects are, however, beyond the scope of this letter. 

The single-particle Hamiltonian ${\cal H}_0(\blr)$ can be diagonalized 
with the eigenvalues $\zeta_N = \omega_c(N+1/2) - \mu$ 
and the eigenfunctions~\cite{Scharnberg} 
\begin{multline}
  \phi_N (\blr|\blq) 
  = 
  \pi^{1/4} \sum_{m=-\infty}^\infty 
  \exp \Big[ \ui \frac{\sqrt{\pi}m}{\lambda}(y- \lambda^2 q_x)+ \ui q_y y \Big] \\
  \times \varphi_N 
  \big( {\frac{x+ (\frac{\sqrt{\pi}m}{\lambda}+ q_y) \lambda^2}{\lambda}} \big), 
  \label{Eq:LL}
\end{multline}
where 
$\omega_c = {|e|H}/{m^*}$, 
$\varphi_N(x) =  \frac{1}{\sqrt{2^N N! \sqrt{\pi} }} H_N(x) e^{-\frac{1}{2}x^2}$, 
$\lambda= (|e|H)^{-1/2}$, 
$H_N(x)$ is the N$^{th}$ Hermite polynomial, and 
$\blq$ represents the quasi-momentum 
in the magnetic sublattices~\cite{Bychkov}. 
We have chosen a rectangular lattice~\cite{Vavilov} 
with edges $a_x = a$ and $a_y = 2a$ ($a= \sqrt{\pi} \lambda$). We introduce 
the magnetic sublattice representation 
\begin{multline} 
G(\blr_1,\blr_2 ;\ui \en) \\
=\sum_{N_1=0 \atop N_2=0}^\infty \int_{\blq_1 \atop \blq_2} 
\phi_{N_1}(\blr_1|\blq_1) 
G_{N_1,N_2}(\blq_1,\blq_2; \ui \en) \phi^*_{N_2}(\blr_2|\blq_2), 
\label{Eq:MSR} 
\end{multline}
with the shorthand notation 
$\int_{\blq}= \int_{-\pi/a}^{\pi/a} \frac{d q_x}{ 2 \pi} 
\int_{-\pi/2a}^{\pi/2a} \frac{d q_y}{ 2 \pi}$. 
As in ref.~\citen{Vavilov}, 
we assume a square vortex lattice, 
and  neglect, for the moment, the impurity potential for simplicity. 

Using the magnetic sublattice representation, 
the normal component of $\widehat{G}$ can be obtained from 
eq.~(\ref{Eq:Gorkov}) as follows: 
\begin{align}
  G_{N_1, N_2}(\blq_1,\blq_2; \ui \en) 
  &= 
  G^0_{N_1}(\ui \en) (2 \pi)^2 \delta(\blq_1, \blq_2) \delta_{N_1,N_2} 
  \qquad \qquad \nonumber \\ 
  & \quad - G^0_{N_1}(\ui \en) \sum_{N_4=0}^{\infty} 
  \Sigma_{N_1,N_4}(\blq_1; \ui \en) \nonumber \\
  &\quad \times G_{N_4, N_2}(\blq_1,\blq_2; \ui \en), \label{Eq:Gorkov2} \\
  \Sigma_{N_1,N_4}(\blq_1; \ui \en) 
    &= 
    \sum_{N_3=0}^{\infty} \Delta_{N_1,N_3}(\blq_1) 
    \Delta^*_{N_4,N_3}(\blq_1) 
    G^0_{N_3}(-\ui \en), 
  \label{Eq:Sigma} 
  \end{align}
where $G^0_N(\ui \en)= (\ui \en - \zeta_N)^{-1} $ 
is the normal state Green's function, 
and $\Delta_{N_1,N_2}(\blq)$ is given by~\cite{Vavilov} 
\begin{multline}
  \Delta_{N_1,N_2}(\blq) 
  = (-)^{N_2} |\Delta| 
  \Big({ \sqrt{2 \pi} \frac{(N_1+N_2)!}{2^{N_1+N_2+1}N_1! N_2!} }\Big)^{1/2} \\
  \times 
  \sum_{p=-\infty}^\infty e^{2 \ui p q_x a } \varphi_{N_1+N_2} 
  \big( \sqrt{2}(q_y \lambda + \frac{\pi \lambda p}{a} ) \big). 
  \label{Eq:DeltaNN}
\end{multline}
Based on ref.~\citen{Brandt}, 
we replace $\Sigma_{N_1,N_4}(\blq_1; \ui \en)$ in eq.~(\ref{Eq:Gorkov2}) 
by its average over $\blq_1$, 
because $\Delta_{N_1,N_3}(\blq_1)$ appears here in a gauge-invariant manner as 
$|\Delta_{N_1,N_3}(\blq_1)|^2$ and 
the $\blq_1$-dependence has a small effect near $H_{c2}$. 
Note that the $\blq_1$-average in the magnetic sublattice representation 
corresponds to the spatial average 
in the real-space representation. 
Now, we obtain 
\begin{equation}
  G_{N_1, N_2}(\blq_1,\blq_2; \ui \en) = 
  (2 \pi)^2 \delta (\blq_1 - \blq_2) \delta_{N_1,N_2} G^B_{N_1}(\ui \en), 
  \label{Eq:GB2} 
\end{equation} 
where 
$G^B_{N_1}(\ui \en) = 
\big( G_{N_1}^0 (\ui \en )^{-1} - {\Sigma^B_{N_1}(\ui \en )} \big)^{-1} $, 
$\Sigma^B_{N_1}(\ui \en ) = 
  \sum_{N_3=0}^\infty  
  \frac{-|\Delta|^2}{\sqrt{4 \pi N_{\rm F}}} 
  \exp \left(-\frac{(N_1-N_3)^2}{4 N_{\rm F}}\right)
  G_{N_3}^0 ( -\ui \en )$, 
and we have used the Gaussian approximation~\cite{Gruenberg} 
in eq.~(\ref{Eq:DeltaNN}) 
because we are interested in the situation $N_{\rm F} = \mu/\omega_c \gg 1$. 

The Green's function~(\ref{Eq:GB2}) is a generalization of the 
Brandt-Pesch-Tewordt Green's function~\cite{Brandt} to a case beyond the 
quasiclassical approximation, 
and it allows us to study the effect of the non-zero cyclotron frequency $\omega_c$. 
In our approximation, $F$ and $ \Delta $ include the off-diagonal elements 
in the Landau level space spanned by eq.~(\ref{Eq:LL}). 
This is important, since 
the off-diagonal matrix elements incorporate 
the phase coherence of the Cooper pairs. 
In the limit of small $\omega_c \tau$, 
our approximation recovers the Green's function of ref.~\citen{Brandt} 
and the $H_{c2}$-line of ref.~\citen{Helfand}. 
This is in contrast to other approximations 
that lead to a reentrant behavior for $ H_{c2}$~\cite{Dukan,Tesanovic}. 
In this respect, 
we believe that our treatment is  adequate for 
the qualitative description of URu$_2$Si$_2$. 
Finally, the effect of impurities on $G^B_N(\ui \en)$ is included~\cite{Vavilov} 
by the replacement 
$\ui \en \to \ui \en -\frac{\omega_c}{2 \pi \tau} \sum_{N_3=0}^{\infty} 
G^B_{N_3}(\ui \en)$, 
if we neglect the anomalous self-energy~\cite{Vavilov} 
assuming the ultraclean limit. 

Inserting the approximate single-particle Green's function 
into the Kubo formula, 
the thermal conductivity $\kappa_{i j}$ of a clean type-II superconductor 
in the low temperature limit is given by~\cite{Houghton,Yoshioka} 
\begin{multline} 
  \frac{\kappa_{yy} + \ui \kappa_{yx}}{T}  
  =  
  \frac{\mu}{6} \int_{-\infty}^\infty
  d \zeta_N 
  G_N^{B(R)} G_{N+1}^{B(A)} 
  \Big\{ 1 - \big( \frac{ \frac{|\Delta|^2}{v_{\rm F}/\Lambda}}
  {\omega_c+ \ui \frac{g_{\rm av}}{\tau}} \big) \\
  \times  
  \Big[ 
  I \big( \frac{ -\ui \frac{g_{\rm av}}{2 \tau} 
    - \zeta_{N+1} }{v_{\rm F}/\Lambda } \big) 
    -
    I \big( \frac{ \ui \frac{g_{\rm av}}{2 \tau} -\zeta_N }{v_{\rm F}/\Lambda } \big) 
  \Big] \Big\}, 
  \label{Eq:kappa}  
\end{multline} 
where $\Lambda= \lambda/\sqrt{2}$, $v_{\rm F}$ is the Fermi velocity, 
and $I(z)= \frac{1}{\sqrt{\pi}} \int_{- \infty}^{\infty} dt \frac{e^{-t^2}}{z-t} $. 
The retarded ($R$) and advanced ($A$) 
Green's functions at zero energy are given by 
\begin{equation} 
  G_N^{B({R/A})} = \Big(
  \pm \ui \frac{g_{\rm av}}{2 \tau} - \zeta_N
    -\frac{|\Delta |^2}{v_{\rm F}/\Lambda} 
    I \big( \frac{ \pm \ui \frac{g_{\rm av}}{2 \tau} + \zeta_N}{v_{\rm F}/\Lambda} \big)
    \Big)^{-1}.
\end{equation}
The sum over Landau level index $N$ is replaced by the integral over $\zeta_N$, 
which is obtained by transforming the $N$-summation 
by the Poisson summation formula, 
and only the non-oscillatory terms are retained. 
The density of states at the Fermi surface, $g_{\rm av}$, 
is determined self-consistently~\cite{Brandt}.  

We now consider a compensated superconductor by introducing 
the pairing interaction 
$\widehat{V}_{\rm pair}= 
\left( {V^{\rm (e,e)}, \atop V^{\rm (h,e)},} 
{V^{\rm (e,h)} \atop V^{\rm (h,h)}} \right)$, 
where ${V^{\rm (\alpha,\beta)}}$ denotes the pair scattering amplitude between 
bands $\alpha$ and $\beta$. 
Neglecting the interband impurity scattering,  
the thermal conductivity of such a superconductor is given by 
$\kappa_{ij} = \kappa_{ij}^{\rm (e)} + \kappa_{ij}^{\rm (h)} $, 
where the contributions of the electrons $\kappa_{ij}^{\rm (e)}$ 
is given by eq.~(\ref{Eq:kappa}), and that of the
holes $\kappa_{ij}^{\rm (h)}$ is given by eq.~(\ref{Eq:kappa}) 
after replacing $\zeta_N \to - \zeta_N$ in the integrand. 
We fix the length scale to the coherence length 
of the electron band, $\xi_0^{\rm (e)}= v^{\rm (e)}_{\rm F}/2 \pi T_c^{\rm (e)}$, 
assuming that the electron band dominates superconductivity~\cite{Kasahara}. 
Then, we specify the following parameters: 
$k_{\rm F}^{\rm (e)} \xi_0^{\rm (e)}$ 
is proportional to the area of the electron Fermi surface; 
$k_{\rm F}^{\rm (h)}/k_{\rm F}^{\rm (e)} = \sqrt{n^{\rm (h)}/n^{\rm (e)}}$ 
measures the degree of compensation; 
$l_{\rm imp}^{\rm (e)}/\xi_0^{\rm (e)}$ and 
$l_{\rm imp}^{\rm (h)}/l_{\rm imp}^{\rm (e)}$ introduce 
the mean free paths of both carriers; 
$\Delta_0^{\rm (h)}/\Delta_0^{\rm (e)}$ 
(or equivalently $\widehat{V}_{\rm pair}$) 
denotes the ratio of zero-field gaps. 
We keep 
$V^{\rm (e,h)}, V^{\rm (h,e)} \neq 0$, 
which results in 
the field dependence 
$\Delta^{\rm (e/h)}(H) = \Delta_0^{\rm (e/h)} \sqrt{1-H/H_{c2}}$ near $H_{c2}$.

\begin{figure}[t] 
\scalebox{0.5}[0.5]{\includegraphics{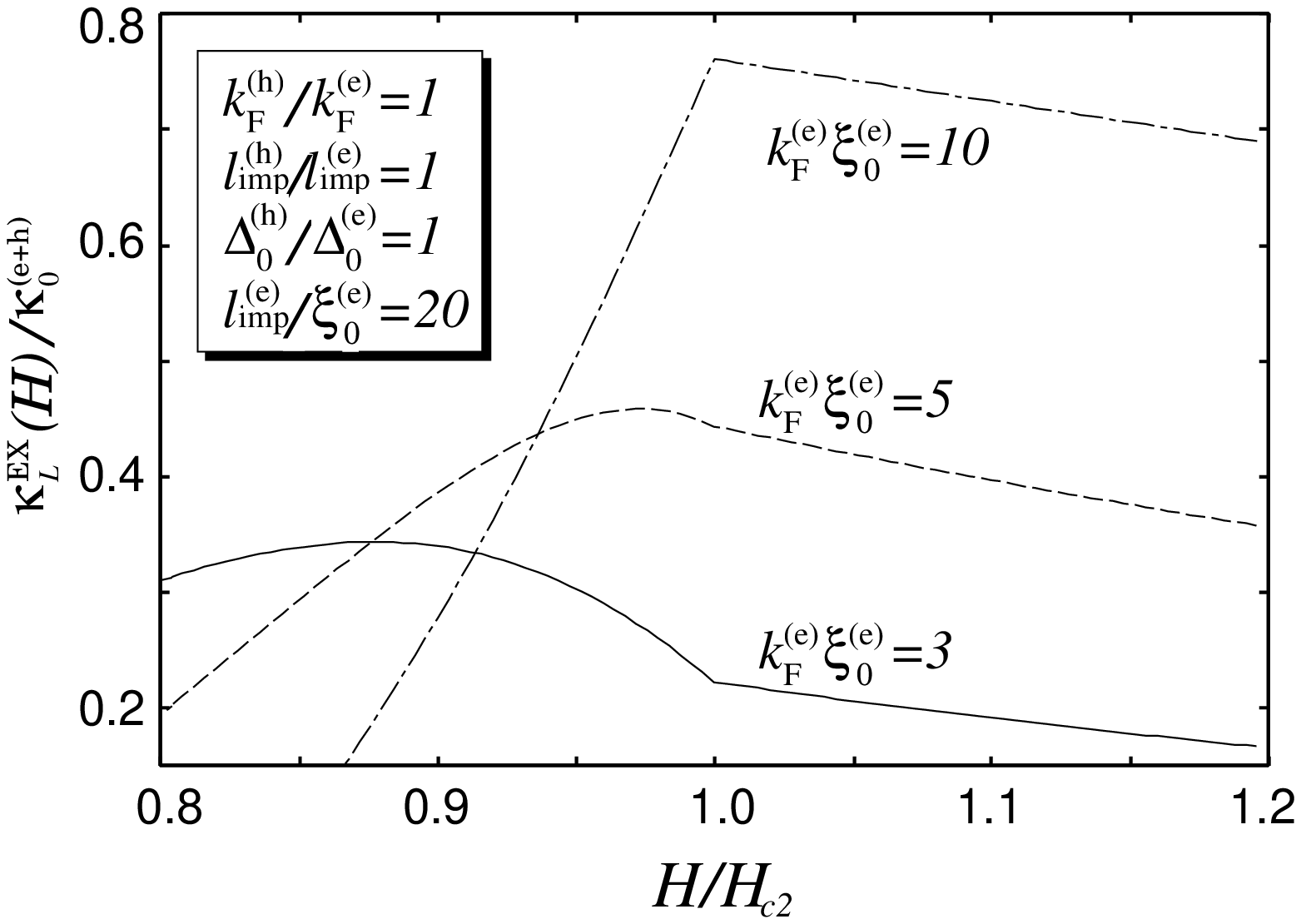}} 
\caption{Magnetic-field dependences of thermal conductivity. 
  Here, $\kappa_0^{\rm (e+h)}= (\frac{\pi}{3}) T 
  ( k_{\rm F}^{\rm (e)} l_{\rm imp}^{\rm (e)}
  + k_{\rm F}^{\rm (h)} l_{\rm imp}^{\rm (h)})$ is the normal state 
  thermal conductivity at zero magnetic field. 
  $\widehat{V}_{\rm pair}= V^{\rm (e,e)}
  \left( {1, \atop 0.1,} {0.1 \atop 1} \right)$ was used. 
} 
\label{Fig:kappa01}
\end{figure}

In the comparison between the theoretical and experimental results 
for  thermal conductivity in a superconductor with 
moderate strength of $\omega_c \tau$, it is important to notice that
experimentally the thermal resistivity tensor $\widehat{\kappa}^{-1}$ 
is measured, and not the thermal conductivity tensor 
$\widehat{\kappa}$~\cite{Vorontsov}. 
Thus, the experimental longitudinal thermal conductivity $\kappa_L^{\rm EX}$ 
is given by 
\begin{equation}
  \kappa_L^{\rm EX} = \kappa_{xx} + \frac{\kappa_{xy}^2}{\kappa_{xx}}, 
 \label{Eq:kappaEX} 
\end{equation}
where we 
used the relations $\kappa_{yy}=\kappa_{xx}$ and 
$\kappa_{yx}= - \kappa_{xy}$. 

First, we address the case of perfect compensation in the normal 
and superconducting states, 
i.e., the states in which the relations 
$k_{\rm F}^{\rm (h)}/k_{\rm F}^{\rm (e)}= 1$,  
$l_{\rm imp}^{\rm (h)}/l_{\rm imp}^{\rm (e)}=1$, and 
$\Delta_0^{\rm (h)}/\Delta_0^{\rm (e)}=1$ hold. 
Figure~\ref{Fig:kappa01} shows the magnetic-field dependences of 
$\kappa_L^{\rm EX}$ for several values of $k_{\rm F}^{\rm (e)} \xi_0^{\rm (e)} $. 
For a large value of $k_{\rm F}^{\rm (e)} \xi_0^{\rm (e)}$ ($=10$), 
the calculated $\kappa_L^{\rm EX}$ decreases for fields below $H_{c2}$. 
This is consistent with the well-known behavior~\cite{Maki,Houghton,Adachi} 
found in quasiclassical calculations 
($k_{\rm F}^{\rm (e)} \xi_0^{\rm (e)} \to \infty$). 
However, as the carrier density (or $k_{\rm F}^{\rm (e)} \xi_0^{\rm (e)}$) 
is reduced, a new property appears. 
On lowering the magnetic field, 
$\kappa_L^{\rm EX}$ initially increases below $H_{c2}$ 
and then decreases, 
forming a hump structure. 
In case of a single carrier-type without electron-hole compensation, 
$\kappa_L^{\rm EX}$ does not show this kind of hump structure 
even for small values of $k_{\rm F}^{\rm (e)} \xi_0^{\rm (e)}$ 
because of the second term in eq.~(\ref{Eq:kappaEX}). 

\begin{figure}[t]
  \scalebox{0.5}[0.5]{\includegraphics{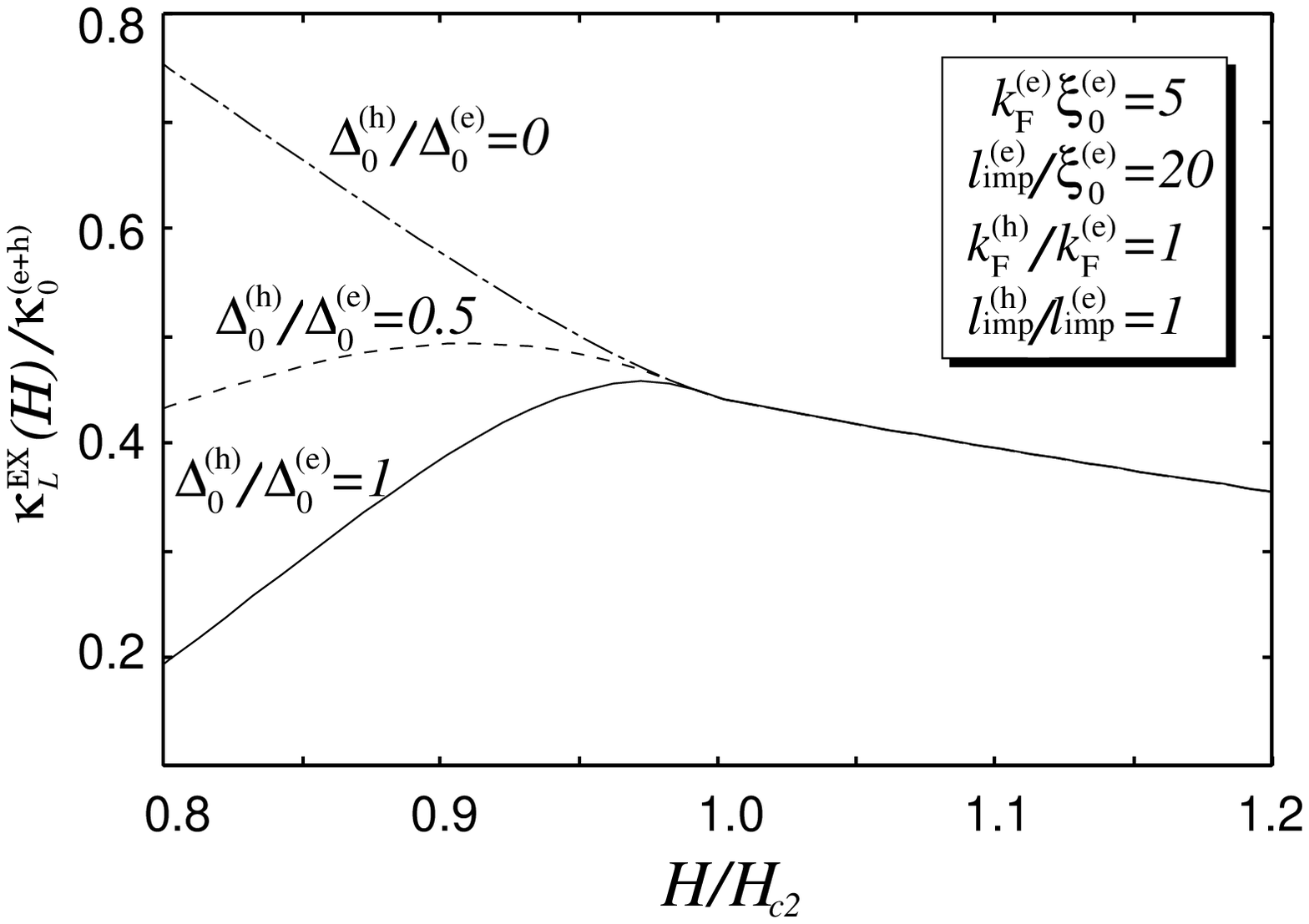}}
  \caption{Enhancement of the thermal conductivity by the 
    deviation from perfect compensation in the superconducting state. 
    $\widehat{V}_{\rm pair}= V^{\rm (e,e)}
    \left( {1, \atop 0,} {0 \atop 0} \right)$, 
    $V^{\rm (e,e)}
    \left( {1, \atop 0.31,} {0.31 \atop 0.5} \right)$, 
    $V^{\rm (e,e)}
    \left( {1, \atop 0.1,} {0.1 \atop 1} \right)$ 
    from top to bottom. 
  } 
    \label{Fig:kappa02} 
\end{figure}

The hump structure of $\kappa^{\rm EX}_{L}$ below $H_{c2}$ 
becomes even more pronounced if the sizes of the energy gaps are different, 
i.e., $\Delta_0^{\rm (h)}/\Delta_0^{\rm (e)} \neq 1$. 
In this case, the second term in eq.~(\ref{Eq:kappaEX}) 
becomes nonzero below $H_{c2}$ and enhances the size of the hump. 
This is actually seen in Fig.~\ref{Fig:kappa02} where  
$\kappa_L^{\rm EX}$ is depicted as a function of magnetic field 
for different ratios of the two gaps. 
With a decreasing ratio $\Delta_0^{\rm (h)}/\Delta_0^{\rm (e)}$, 
the increase in $\kappa_L^{\rm EX}$ below $H_{c2}$ is enhanced. 
Evidently, the violation of compensation in the 
superconducting phase enhances the hump structure 
below $H_{c2}$.

\begin{figure}[t]
  \scalebox{0.5}[0.5]{\includegraphics{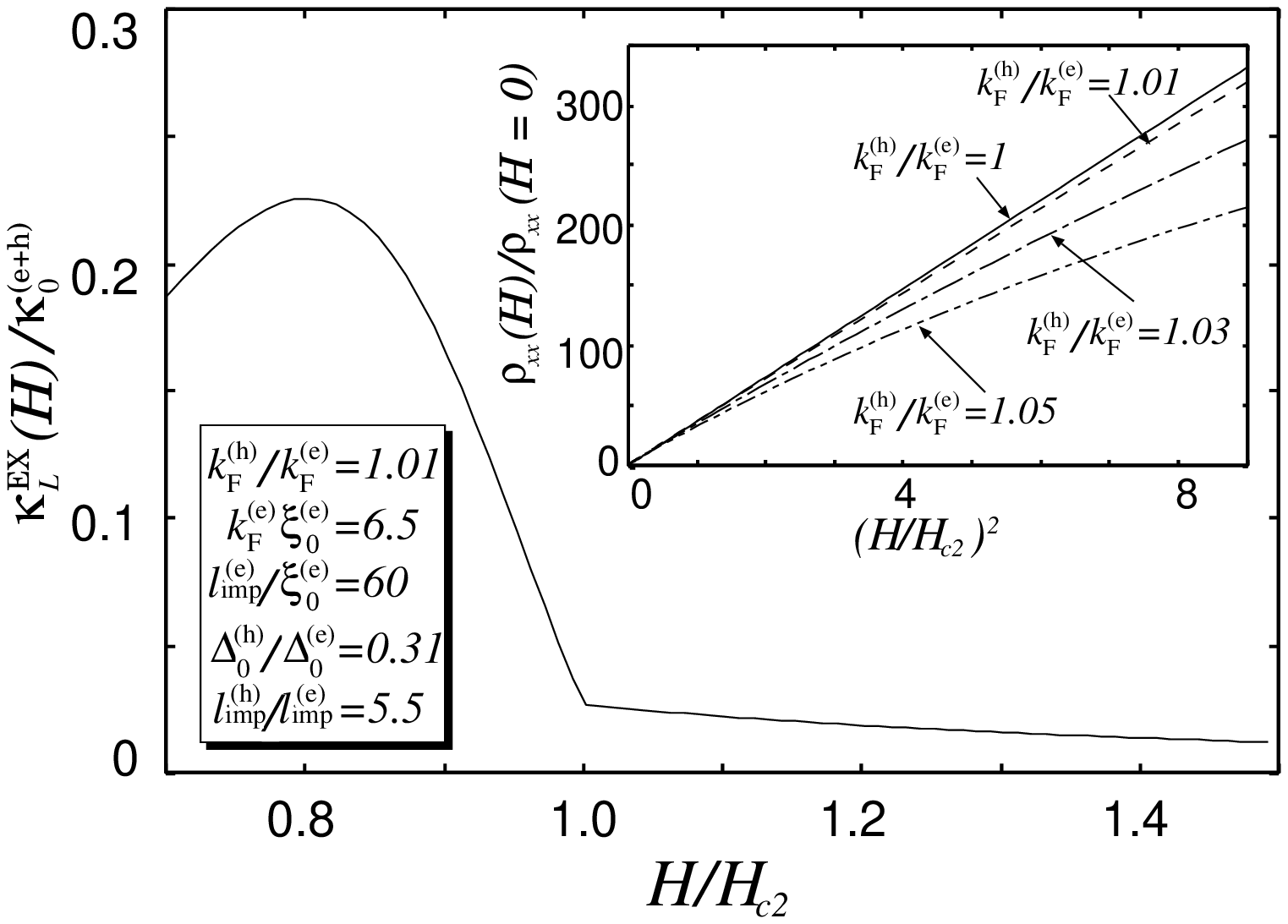}}
  \caption{Main panel: 
    Calculated thermal conductivity for estimated parameters of URu$_2$Si$_2$. 
    $\widehat{V}_{\rm pair}= V^{\rm (e,e)}
    \left( {1, \atop 0.35,} {0.35 \atop 0} \right)$. 
    Inset: Magneto-resistance as a function of $H^2$ 
    for several values of $k_{\rm F}^{\rm (h)}/k_{\rm F}^{\rm (e)}$.} 
  \label{Fig:kappa03}
\end{figure}

Finally, we compare our results with the mixed-state 
thermal conductivity of URu$_2$Si$_2$.  
Note that our analysis yields only a qualitative understanding 
because we employ a simple quasiparticle picture 
with a two-dimensional isotropic Fermi surface and 
neglect the paramagnetic effect as well as the correlation effects.
We fix the parameters as follows. 
From the values $H_{c2} \approx 2.8$ T for ${\bm H} \parallel$ c 
together with $k_{\rm F} \approx 1.1$ nm$^{-1}$~\cite{Behnia1}, 
and taking into account the fact that there is a substantial 
paramagnetic effect~\cite{Brison} in this material, 
we have a rough estimate $k_{\rm F} \xi_0 \simeq 6.5$. 
For the mean free path, 
we use $l_{\rm imp}^{\rm (e)} / \xi_0^{\rm (e)} =60 $ and 
$l_{\rm imp}^{\rm (h)} / l_{\rm imp}^{\rm (e)}=5.5$ 
in order to reproduce the observed magneto-resistance 
($\Delta \rho_{xx}(10{\rm T}) /\rho_{xx}(0{\rm T}) \approx 300 $) 
with a nearly perfect $H^2$-dependence 
(see the inset of Fig.~\ref{Fig:kappa03}). 
This choice of parameters is consistent with the scenario~\cite{Kasahara} 
in which the Hall effect is dominated by the light hole band. 
Further, the positive Hall coefficient 
and the measured value~\cite{Shibauchi}  
$\rho_{xy}(10{\rm T})/\rho_{xx}(0{\rm T}) \approx 50$
gives an estimate $k_{\rm F}^{\rm (h)}/k_{\rm F}^{\rm (e)}= 1.01$. 
Finally, we have assumed that a larger gap is formed 
in the electron band following the discussion in ref.~\citen{Kasahara}. 
For the gap ratio, we use 
$\Delta_0^{\rm (h)}/\Delta_0^{\rm (e)}= 0.31$. 

The main panel of Fig.~\ref{Fig:kappa03} shows the calculated thermal 
conductivity of URu$_2$Si$_2$ as a function of the magnetic field. 
The peculiar magnetic-field dependence 
of the thermal conductivity with the hump structure below $H_{c2}$ is 
reproduced, with its size being 
slightly larger than the measured one in our calculation~\cite{Kasahara}.
The neglected interband impurity scattering 
would reduce the size of the hump. 
It should be noted that 
while the size of the hump is modified by changing the parameters, 
the appearance of the hump itself is robust for parameters reproducing 
the magneto-resistance data.~\cite{Kasahara} 
In the experiment, the thermal conductivity increases sharply 
below $H_{c2}$, which most probably results from a paramagnetic limiting effect 
in this material, which we neglect here. In our model case, the hump develops more
gently. 
To treat these other effects is beyond the scope of our study and therefore 
we leave it for future studies. 

\begin{figure}[t]
\scalebox{0.5}[0.5]{\includegraphics{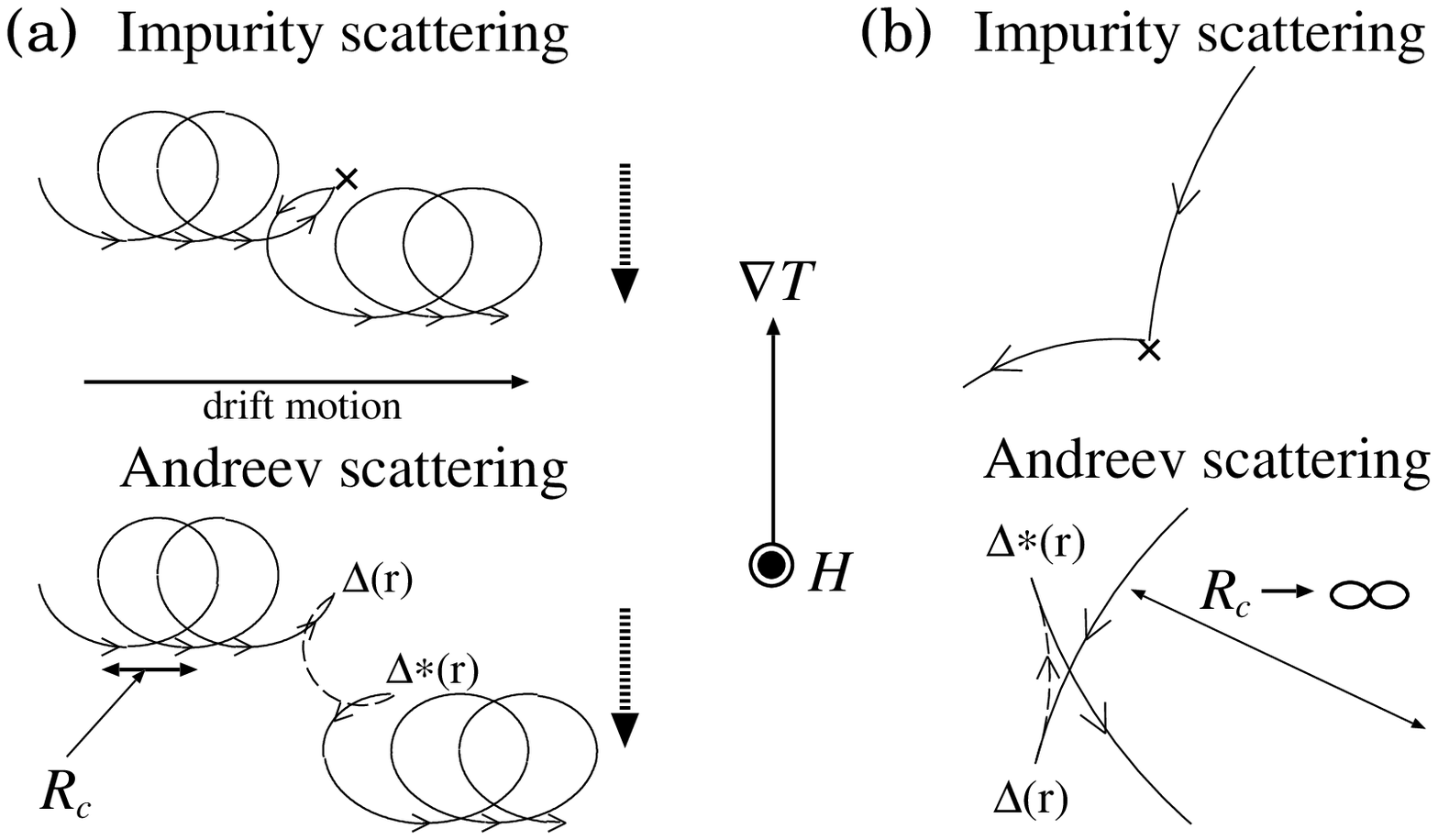}} 
\caption{Schematics of the quasiparticle trajectory 
  where $R_c= \lambda^2 k_{\rm F} $ is the radius of the 
  cyclotron motion of quasiparticles; 
  (a) high field ($\omega_c \tau \gg 1$), 
  (b) low field ($\omega_c \tau \ll 1$). 
}
\label{Fig:cartoon} 
\end{figure}

The special scattering features provide some physical insight into this behavior. 
For the superclean limit $\omega_c \tau \gg 1$,~\cite{com1} 
the heat transport along $- {\bm \nabla} T$ by quasiparticles 
is only possible through scattering (see Fig.~\ref{Fig:cartoon}(a)), as 
otherwise the quasiparticles would experience only the drift motion perpendicular 
to  $- {\bm \nabla} T$. 
In the normal state at low temperatures, this corresponds mainly to 
impurity scattering. In the superconducting mixed phase, an additional 
contribution is derived from Andreev scattering.~\cite{Houghton,com3} 
Hence, at high magnetic fields, 
the Andreev scattering tends to {\it increase} the thermal conductivity 
below $H_{c2}$. 
At lower fields (smaller $\omega_c \tau$), 
the Andreev scattering plays a different role. The particle and hole trajectories 
tend to retrace (Fig.~\ref{Fig:cartoon}(b)). 
Because both quasiparticle types carry heat, 
a compensation occurs, which decreases the heat current parallel to $- {\bm \nabla} T$ 
and $\kappa_{L}^{\rm EX}$ decreases at lower fields. 
The overall behavior leads to the characteristic hump feature. 
Note that this picture of the single-carrier $\kappa_{xx}$ is applicable to 
$\kappa_{L}^{\rm EX}$ because electron-hole compensation leads to 
the suppression of the second term in eq.~(\ref{Eq:kappaEX}). 
In the ordinary case without the compensation, this picture cannot be applied 
to $\kappa_{L}^{\rm EX}$ because the second term in eq.~(\ref{Eq:kappaEX}) 
makes an important contribution to $\kappa_{L}^{\rm EX}$ at high fields, 
veiling the hump below $H_{c2}$.  

In conclusion, 
we have examined the effects of the low carrier density and 
the electron-hole compensation on mixed-state thermal transport. 
The interplay of these two effects 
leads to the appearance of a hump structure of 
$\kappa_{xx}(H)$ below $H_{c2}$, as observed in URu$_2$Si$_2$, 
whose electronic states incorporate the two features in the hidden-order phase. 
Our study provides a natural explanation for the 
unusual magnetic-field dependence of $\kappa_{xx}(H)$ near $H_{c2}$. 
Moreover, it demonstrates that 
URu$_2$Si$_2$ provides a rare chance to examine the physics of the mixed phase 
in the superclean limit, resulting in intriguing magneto-transport phenomena.

We are grateful to Y. Matsuda, T. Shibauchi, Y. Kasahara and N. Hayashi 
for insightful discussions. 
This study was financially supported through a fellowship of the 
Japan Society for the Promotion of Science and the NCCR MaNEP 
of the Swiss National foundation.

\bibliography{basename of .bib file}

\begin{thebibliography}{99}
\bibitem{Xu}
  Z. A. Xu, N. P. Ong, Y. Wang, T. Kakeshita, and S. Uchida: 
  Nature {\bf 406} (2000) 486. 
\bibitem{Bel1} 
  R. Bel, K. Behnia, and H. Berger: 
  Phys. Rev. Lett. {\bf 91} (2003) 066602. 
\bibitem{Behnia1}
  K. Behnia, M. -A. M\'{e}asson, and Y. Kopelevich: 
  Phys. Rev. Lett. {\bf 98} (2007) 076603. 
\bibitem{Kasahara}
  Y. Kasahara, T. Iwasawa, H. Shishido, T. Shibauchi, 
  K. Behnia, Y. Haga, T. D. Matsuda, Y. Onuki, M. Sigrist, and Y. Matsuda: 
  Phys. Rev. Lett. {\bf 99} (2007) 116402. 
\bibitem{com2}
  Since the inelastic scattering rate usually decreases as $T^\alpha (\alpha > 0)$ 
  at low temperatures, 
  it is unlikely that this result in the low-temperature limit 
  is caused by the suppression of the e-e scattering due to the 
  opening of the superconducting gap, as discussed in ref.~\citen{Hirschfeld}.  
\bibitem{Hirschfeld}
  P. J. Hirschfeld and W. O. Putikka: 
  Phys. Rev. Lett. {\bf 77} (1996) 3909; 
  H. Hara and H. Kontani, J. Phys. Soc. Jpn. {\bf 76} (2007) 073705. 
\bibitem{Maki} 
  K. Maki: Phys. Rev. {\bf 158} (1967) 397. 
\bibitem{Houghton} 
  A. Houghton and K. Maki: 
  Phys. Rev. B {\bf 4} (1971) 843. 
\bibitem{Adachi} 
  H. Adachi, P. Miranovi\'{c}, M. Ichioka, and K. Machida: 
  J. Phys. Soc. Jpn. {\bf 76} (2007) 064708. 
\bibitem{Palstra}
  T. T. M. Palstra, A. A. Menovsky, J. van den Berg, 
  A. J. Dirkmaat, P. H. Kes, G. J. Nieuwenhuys, and J. A. Mydosh: 
  Phys. Rev. Lett. {\bf 55} (1985) 2727. 
\bibitem{Maple} 
  M. B. Maple, J. W. Chen, Y. Dalichaouch, T. Kohara, C. Rossel, 
  M. S. Torikachvili, M. W. McElfresh, and J. D. Thompson: 
  Phys. Rev. Lett. {\bf 56} (1986) 185. 
\bibitem{Baumann} 
  W. Schlabitz, J. Baumann, B. Pollit, U. Rauchschwalbe, H. M. Mayer, 
  U. Ahlheim, and C. D. Bredl: 
  Z. Phys. B {\bf 62} (1986) 171. 
\bibitem{Schoenes}
  J. Schoenes, C. Sch\"{o}nenberger, J. J. M. Franse, and 
  A. A. Menovsky: 
  Phys. Rev. B {\bf 35} (1987) 5375. 
\bibitem{Bel2}
  R. Bel, H. Jin, K. Behnia, J. Flouquet, and P. Lejay: 
  Phys. Rev. B {\bf 70} (2004) 220501(R). 
\bibitem{Behnia2}
  K. Behnia, R. Bel, Y. Kasahara, Y. Nakajima, H. Jin, H. Aubin, K. Izawa, 
  Y. Matsuda, J. Flouquet, Y. Haga, Y. Onuki, and P. Lejay: 
  Phys. Rev. Lett. {\bf 94} (2005) 156405. 
\bibitem{Brandt} 
  U. Brandt, W. Pesch, and L. Tewordt: 
  Z. Phys. {\bf 201} (1967) 209. 
\bibitem{Vavilov}
  M. G. Vavilov and V. P. Mineev: 
  Zh. Eksp. Teor. Fiz. {\bf 112} (1997) 1873 
  [Sov. Phys. JETP {\bf 85} (1997) 1024]. 
\bibitem{Ohkuni} 
  H. Ohkuni, Y. Inada, Y. Tokiwa, K. Sakurai, R. Settai, T. Honma, Y. Haga, 
  E. Yamamoto, Y. Onuki, H. Yamagami, S. Takahashi, and T. Yanagisawa: 
  Philos. Mag. B {\bf 79} (1999) 1045. 
\bibitem{Brison} 
  J. P. Brison, N. Keller, P. Lejay, A. Huxley, L. Schmidt, A. Buzdin, 
  N. R. Bernhoeft, I. Mineev, A. N. Stepanov, J. Flouquet, D. Jaccard, 
  S. R. Julian, and G. G. Lonzarich: 
  Physica C {\bf 199-200} (1994) 70. 
\bibitem{Scharnberg} 
  K. Scharnberg: 
  J. Low. Temp. Phys. {\bf 6} (1972) 51. 
\bibitem{Bychkov} 
  Yu. A. Bychkov and E. I. Rashba:
  Zh. Eksp. Teor. Fiz. {\bf 85} (1983) 1826 
  [Sov. Phys. JETP {\bf 58} (1983) 1062]. 
\bibitem{Gruenberg} 
  L. W. Gruenberg and L. Gunther: 
  Phys. Rev. {\bf 176} (1968) 606. 
\bibitem{Helfand}
  E. Helfand and N. R. Werhamer: 
  Phys. Rev. {\bf 147} (1966) 288. 
\bibitem{Dukan} 
  S. Dukan and Z. Te\v{s}anovi\'{c}: 
  Phy. Rev. B {\bf 56} (1997) 838. 
\bibitem{Tesanovic} 
  Z. Te\v{s}anovi\'{c}, M. Rasolt, and L. Xing: 
  Phys. Rev. Lett. {\bf 63} (1989) 2425. 
\bibitem{Yoshioka} 
  D. Yoshioka: 
  Phys. Rev. B {\bf 27} (1983) 3637. 
\bibitem{Vorontsov}
  A. Vorontsov and I. Vekhter: 
  Phys. Rev. B {\bf 75} (2007) 224502. 
\bibitem{Shibauchi}
  T. Shibauchi: private communication. 
\bibitem{com1} 
  The condition $\omega_c \tau \gg 1 $ at $H \approx H_{c2}$ corresponds 
  to ${\Delta^2 \tau}/{\mu} \gg 1$ at $H \ll H_{c2} $, which 
  is usually called the superclean limit. 
\bibitem{com3}
  A picture of Andreev scattering by vorteces is valid under the condition 
  $(\frac{|\Delta|}{v_{\rm F}/\Lambda})^2 < \frac{\Lambda}{l_{\rm imp}}
  [1+ (\omega_c \tau)^2]$. 
\end{thebibliography}

\end{document}